\documentclass[a4paper]{article}

\usepackage{graphicx}
\begin{document}
\tolerance=10000
\null
\hyphenation{visco-elastic visco-elasticity}
\font\title=cmbx10 scaled \magstep2
\font\text=cmr10 at 12 truept
\font\refs=cmr10 at 11 truept
\font\note=cmr10 at 10 truept
\font\bf=cmbx10
\font\it=cmti10
\font\sc=cmcsc10
\font\sl=cmsl10
\def\pni{\par \noindent}
\def\vsh{\smallskip}
\def\vs{\medskip}
\def\vvs{\bigskip}
\def\vvvs{\bigskip\medskip} 
\def\vsp{\vsh \par}
\def\vsn{\vsh\pni}
\def\cen{\centerline}
\def\ra{\item{a)\ }} \def\rb{\item{b)\ }}   \def\rc{\item{c)\ }}
\def\eg{{\it e.g.}\ } \def\ie{{\it i.e.}\ }
\def\rl{\vsn}
\def\e{{\rm e}}
\def\exp{{\rm exp}}
\def\ds{\displaystyle}
\def\dis{\displaystyle}
\def\q{\quad}	 \def\qq{\qquad}
\def\lan{\langle}\def\ran{\rangle}
\def\l{\left} \def\r{\right}
\def\lra{\Longleftrightarrow}
\def\d{\partial}
\def\dr{\partial r}  \def\dt{\partial t}
\def\dx{\partial x}   \def\dy{\partial y}  \def\dz{\partial z}
\def\rec#1{{1\over{#1}}}
\def\bar{\tilde}
\def\barr{\widetilde}
\def\epsilons{{\widetilde \epsilon(s)}}
\def\sigmas{{\widetilde \sigma (s)}}
\def\fs{{\widetilde f(s)}}
\def\Js{{\widetilde J(s)}}
\def\Gs{{\widetilde G(s)}}
\def\Fs{{\wiidetilde F(s)}}
 \def\Ls{{\widetilde L(s)}}
\def\L{{\cal L}} 
\def\F{{\cal F}} 
\def\NN{{\rm I\hskip-2pt N}}
\def\RR{\vbox {\hbox to 8.9pt {I\hskip-2.1pt R\hfil}}}
\def\CC{{\rm C\hskip-4.8pt \vrule height 6pt width 12000sp\hskip 5pt}}
\def\II{{\rm I\hskip-2pt I}}
\def\I{{\cal I}}  
\def\D{{\cal D}}  
\def\Gc{{\cal {G}}_c}	\def\Gcs{\barr{\Gc}} 
\def\Gs{{\cal {G}}_s}	\def\Gss{\barr{\Gs}} 
\def\args{(x/ \sqrt{D})\, s^{1/2}}
\def\argsa{(x/ \sqrt{D})\, s^{\beta}}
\def\arg{ x^2/ (4\,D\, t)}
 \setcounter{page}{1}   \thispagestyle{empty}
{\markboth
 {\rm \centerline {F. Mainardi and  M. Tomirotti}}
 {\rm \centerline{Seismic pulse propagation with constant $Q$...\hfill}}
 }
\cen{ANNALI DI GEOFISICA, Vol. 40, No 5, pp. 1311-1328, October 1997}
 \vsh\hrule
  $\null$
\vskip 2.0truecm
\cen{{\title Seismic pulse propagation with constant $Q$}}
\vsh
\cen{{\title and  stable probability   distributions}}

\vskip 1.0 truecm
\cen{Francesco MAINARDI $\null^{(1)}$ and
     Massimo TOMIROTTI	$\null^{(2)}$}

\vsh
\cen{$\null^{(1)}$ Dipartimento di Fisica, Universit\`a di Bologna}
\cen{Via Irnerio 46, I-40126 Bologna, Italy}
\cen{{\it e-mail: francesco.mainardi@unibo.it}}
\vsh
\cen{$\null^{(2)}$ Dipartimento di Ingegneria Idraulica, Ambientale e del
     Rilevamento}
\cen{Politecnico di Milano,
Piazza Leonardo da Vinci 32, I-20133 Milano, Italy}
\cen{{\it e-mail: tom@idra5.iar.polimi.it}}
 \vskip 1.0 truecm
 \noindent {\it This note is dedicated to Professor Michele Caputo
in occasion of his 70-th birthday.
Throughout his intensive and outstanding career
Professor Caputo has recognized  the importance  of
the quality factor $Q\, $ and  fractional calculus in seismology,
providing interesting contributions on these topics.}
 
\section*{Abstract}
\vsn
The one-dimensional propagation of seismic waves with constant $Q$ is shown
to  be governed by an evolution equation  of fractional order
in time, which	interpolates the heat equation and the wave equation.
The fundamental solutions for the {Cauchy} and
{Signalling} problems are
expressed
in terms of  entire functions (of Wright type) in the similarity variable
and their behaviours turn out to be intermediate between
those for the limiting	cases of a perfectly  viscous fluid
and a perfectly elastic solid.
In view of the small dissipation
exhibited by the seismic pulses, the nearly elastic limit
is considered. 
Furthermore,
the fundamental solutions  for the {Cauchy} and {Signalling} problems
are shown to
be related to  {stable} probability  distributions
with index of stability  determined by the order of the
fractional time derivative in the evolution  equation.

\vs  \noindent
{\bf Key words:} {\it  Earth anelasticity} -
{\it Quality factor} - {\it Wave propagation} -
{\it Fractional derivatives}  - {\it Stable probability distributions}
\newpage
\section{Introduction}
\vs\vsp
In seismology the problem of wave attenuation
due to anelasticity of the Earth
is described
by the so-called   quality factor $Q\,, $ or, better, by its inverse
$Q^{-1}\, $ (internal friction or  loss tangent),
which is related to  the dissipation of the elastic
energy during the wave propagation. Because of its
great relevance in determining the composition and the mechanical
properties of the Earth,  the problem has  been considered
from different points of view by many researchers.
Without pretending to be exhaustive, we quote (in alphabetic order
of the first author) some  original contributions and reviews
among those which have	attracted our attention,
\eg
Aki and Richards (1980), Ben-Menhahem and Singh (1981),
Caputo (1966, 1967, 1969, 1976, 1979, 1981, 1985, 1996a)
Caputo and Mainardi (1971),
Carcione  {\it et al.} (1988),
Chin (1980),  
Futterman (1962),
Gordon and Nelson (1966),
Jackson and Anderson (1970),
Kanamori and Anderson (1977),
Kang and McMechan (1993)
Kjartansson (1979),
Knopoff (1964),
K\"ornig and M\"uller (1989),
Mitchell (1995),
Murphy (1982),
O'Connel and Budiansky (1978),
Ranalli (1987),
Sabadini {\it et al.}  (1985, 1987),
Savage and O'Neill (1975),
Spencer (1981),
Strick (1967, 1970, 1982, 1984), Strick and  Mainardi (1982),
 Yuen {\it et al.} (1986).
\vsp
It is known that seismic pulse propagation mostly occurs
with a quality	factor $Q$  constant  over a wide
range of frequencies.
As pointed out by Caputo and Mainardi
(1971) and  Caputo (1976),
this factor turns out to be independent of frequency
only in special linear viscoelastic media  for which
the  stress  is proportional to a fractional derivative of the strain,
of order $\nu$ less than one.
Since these media
exhibit a creep compliance depending on time by a power-law
with exponent  $\nu $,
we refer to them  as power-law solids,
according to the notation by Kolsky (1956)
and Pipkin (1972-1986).
\vsp
For the sake of convenience, the generalized operators of
integration and differentiation
of arbitrary order are recalled in the Appendix in the
framework of the so-called Riemann-Liouville {\it Fractional Calculus}.
In this paper we adopt the Caputo definition
for the {\it fractional derivative of order}
$\alpha >0 $ of a causal function $f(t)$ (\ie vanishing for $t<0$),
$$    {d^\alpha \over dt^\alpha }\, f(t) :=
    \left\{
	 \begin{array}{ll}
	{\ds f^{(m)}(t)} \; &\; {\rm if} \q
	   \alpha = m \in \NN\,, \\
      { \ds \rec{\Gamma(m-\alpha )} \, \int_0^t
  {f^{(m)} (\tau )\over (t-\tau )^{\alpha +1-m }} \, d\tau}
  & \;{\rm if} \q  m-1 <\alpha < m\,,   
  \end{array}
  \right.
  \eqno(1.1)$$
where $f^{(m)}(t)$ denotes the	derivative of integer order $m$
and $\Gamma$ is the Gamma function.
\vsp
In Section 2 we derive the general evolution equation
governing the  propagation of uniaxial stress waves,
in the framework of the  dynamical theory
of linear viscoelasticity.
For a power-law solid
the evolution equation is shown to be
of fractional order in time, which is intermediate between the heat
equation and the wave equation.
In fact, denoting  the space and time variables by $x$ and $t$
and  the response field variable by $w(x,t)$,
the evolution equation
will be shown to be
  $${\d^{2\beta } w\over \dt^{2\beta} } =
  D\,{\d^2 w \over \dx^2}  \,,	\qq 2\beta   =2-\nu \,.
\eqno(1.2)$$
The order of the time derivative has been denoted by $2\beta $
for reasons that will appear later.
Since $0<\nu \le 1\,, $ we get	$1/2 \le \beta	< 1\,. $
\vsp
In Section 3
 we  review the analysis of the fractional
evolution equation (1.2) in the  general case
$0 <\beta  < 1\,, $
essentially  based on our works, Mainardi (1994, 1995, 1996a, 1996b).
We first analyse
the two basic boundary-value problems,
referred to as
the {\it Cauchy} problem  and the {\it Signalling} problem,
by   the technique of  the Laplace transform and
we derive  the transformed expressions
of the respective fundamental solutions (the {Green functions}).
Then, we  carry out the inversion of the relevant Laplace transforms
and we outline a {\it reciprocity relation} between the Green functions
in the space-time domain.  In view of this relation
the Green functions can be expressed
in terms  of two interrelated {\it auxiliary functions}  in
the similarity variable $r = |x|/(\sqrt{D}t^\beta )\,.$
These auxiliary
functions can be analytically continued in the whole complex plane
as entire functions of {Wright} type.
\vsp
In Section 4 we show the evolution of the fundamental solutions
for $ 1/2 \le \beta  <1$,  that
can be relevant   in
seismology to simulate the propagation of seismic pulses.
Accounting for the  low dissipation occurring in the Earth, the nearly
elastic limit  must be	considered;
in this case the pulse response becomes a narrow,
sharply peaked function and
the arguments by Pipkin (1972-1986) and Kreis
and Pipkin (1986) must be adopted in order to obtain
an evaluation of the solutions, which is suitable from numerical point of
view.
\vsp
Finally, in Section 5, following Kreis and Pipkin (1986),
 we point out the interesting connection between
the  fundamental solution for the {\it Signalling} problem and the density
of a certain (unilateral) {\it stable} probability distribution.
We note that
this   connection
generalizes  the one known for the standard heat
equation for which the fundamental solution
for the {\it Signalling} problem is
related to the density of the {\it stable} {\it L\'evy} distribution.
Since the above property is expected to provide a further insight
into our evolution equation of fractional order,
the seismic pulse propagation with constant $Q$ assumes
an additional interest
from a mathematical-physical point of view.  
\vvs
\section{Linear Viscoelastic Waves and the
 Fractional Diffusion-Wave
Equation}
 \vs\vsp
According to the elementary one-dimensional theory of linear
viscoelasticity, the medium  is assumed to be homogeneous
(of density $\rho $), semi-infinite or infinite in extent
($0 \le x<+\infty $ or	 $-\infty <x < +\infty$) and
undisturbed for $t<0\,. $
The basic equations are known to be,
see \eg Hunter (1960), Caputo \& Mainardi (1971), Pipkin (1972-1986),
Christensen (1972-1982), Chin (1980), Graffi (1982),
$$ \sigma _x(x,t)  = \rho \, u_{tt}(x,t)\,, \eqno(2.1)$$
$$  \epsilon (x,t) = u_x(x,t)\,, \eqno(2.2)$$
$$  \epsilon (x,t) = [J_0  + \dot J(t) * \,] \,\sigma (x,t)
    \,. \eqno(2.3)$$
Here the suffices $x$ and $t$
denote partial derivation with respect
to space and time respectively,  the dot  ordinary time-derivation,
and the star  integral time-convolution from $0^+$ to $t$.
The following notations have been used:
$\sigma $ for the stress, $\epsilon  $	for  the strain,
$J(t) $ for  the creep compliance (the strain response to a
unit step input of stress); the constant
$J_0 :=  J(0^+) \ge 0$ denotes
 the instantaneous (or glass) compliance.
\vsp
The evolution equation for the {\it response variable}
 $w(x,t)$ (chosen among
the field variables:
the displacement $u$, the stress $\sigma $, the strain $\epsilon $
or the particle velocity $v=u_t$) can be derived through the
application of the Laplace transform to the basic equations.
We use the following notation
for the Laplace transform of a	function  $f(t)\,, $
 locally summable for $t\ge 0\,, $
 $$ {\cal{L}}\, \l\{  f(t) \r\}	:= \int_0^\infty \!\!
  \e^{-st}\, f(t)\, dt = \widetilde f(s)\,, \; s \in \CC\,,$$
and we adopt  the sign $\div$ to denote a Laplace transform pair,
\ie
$ f(t) \div  \widetilde f(s) \,. $
\vsp
We first obtain in the transform domain,
the  second order differential equation 
$$\l[ {d^2\over dx^2 }- \mu ^2(s) \r]\, \widetilde w(x,s)=0\,, \eqno(2.4)
$$ in which
$$\mu (s) := s \, \l[\rho \, s \Js \r]^{1/2} \eqno(2.5)  $$
is real and positive for $s$ real and positive.
As a matter of fact, $\mu (s)$ turns out to be an analytic
function of $s$ over the entire $s$-plane cut along
the negative real axis; the cut  can be limited or unlimited
in accordance with the particular visco\-elastic model assumed.
\vsp
Wave like or diffusion like  character of the evolution equation
 can be drawn from (2.5) by taking into
account the asymptotic representation of the creep compliance
for short times,
  $$
J(t) = J_0  +O(t^\nu )\,, \q {\rm as}\; t \to 0^+\, ,\eqno(2.6)$$
 with $J_0 \ge 0\,,$ and $0<\nu \le 1  \,.$
If $J_0 >0$ then
$$ \lim_{s\to \infty} {\mu (s)\over s} = \sqrt{\rho J_0}
   := {1 \over c}\,,   \eqno(2.7)$$
 we have a wave like behaviour with $c$ as the wave-front velocity;
otherwise ($J_0 =0)$ we have a diffusion like behaviour.
In the	case $J_0>0$ the wave like
evolution equation for $w(x,t)$
can be derived by inverting (2.4-5), using (2.6-7)
and introducing the non dimensional rate of creep
 $$  \psi(t) :=  {1\over J_0}\,  {dJ(t)\over dt} \ge 0\,,
   \q t>0\,. \eqno(2.8)$$
We get
$$  \mu ^2(s) := s^2 [\rho \, s\Js]
= \l({s\over c}\r)^2  [1 + \widetilde \psi(s)]\,, \eqno(2.9)$$
so that the evolution equation turns out to be
$$
 \l\{ 1 + \psi(t) \, *\, \r\}  \, {\d^2 w\over \dt^2}
       = c^2\,{\d^2 w\over \dx^2}
    \,. \eqno(2.10)$$
This is  a generalization of  D'Alembert  wave equation
in that it is an integro-differential equation
where the   convolution integral can be interpreted as a
perturbation term.
This case has been investigated by Buchen and  Mainardi (1975)
and by Mainardi and Turchetti (1975), who have provided
wave-front expansions for the solutions.
\vsp
In the	case $J_0 =0$ we can re-write (2.6)  as
   $$  J(t) = {1\over \rho D} \, {t^\nu  \over \Gamma(\nu +1)}
  +  o\, (t^\nu ) \,,\q {\rm as}\; t \to 0^+\,,    \eqno(2.11)$$
where, for the sake of convenience, we have  introduced
 the positive constant $D$ (whose dimensions
are $L^{2}\, T^{\nu -2}$)
 and  the Gamma function $ \Gamma(\nu +1)\,. $
Then we can introduce
the non-dimensional function  $\phi(t)$  whose
Laplace transform is such that
$$ \mu ^2(s) := s^2 \,[\rho \, s\Js]
  ={s^{2-\nu } \over D} \, [1 +\widetilde\phi(s)]  \,. \eqno(2.12)$$
Using (2.12), the Laplace inversion of (2.4-5)	 yields
$$
 \l\{ 1 + \phi(t) \, *\, \r\}  \, {\d^{2\beta } w \over \dt^{2\beta } }
       = D\,{\d^2 w\over \dx^2}
    \,, \eqno(2.13)$$
where
  $ 2\beta  = 2 - \nu	\,$ so $\,  1/2\le \beta  <1\,.$
Here the time-derivative turns out
to be just the	fractional derivative of order $2\beta$ (in Caputo's
sense), according to the Riemann-Liouville theory of
{\it Fractional Calculus} recalled  in the Appendix.
\vsp
When
the creep compliance satisfies the simple power-law
$$ J(t) =    {1\over \rho D} \, {t^\nu	\over \Gamma(\nu +1)}
\,,\q t > 0\,,	  \eqno(2.14)$$
we obtain $\phi(t) \equiv 0\,, $
and  the evolution equation (2.13) simply reduces to  (1.2).
As pointed out by  Caputo and Mainardi (1971),	 the creep law
(2.14)
is provided by	viscoelastic  models whose   stress-strain relation
(2.3)  can be simply expressed	by a fractional derivative
of order $\nu \,. $ In the present notation this stress-strain
relation reads
$$ \sigma  = { \rho D}\,{d^\nu	 \over dt^\nu }\,\epsilon
 \,, \q 0<\nu \le 1\,. \eqno(2.15)$$
For $\nu =1$ the Newton law for a viscous fluid is recovered
from (2.15) where $D$ now represents the kinematic  viscosity;
in this case, since $\beta  =1/2$ in (1.2),
the classical diffusion equation (or heat equation) holds for $w(x,t)\,.$
When $ 0<\nu <1$ the evolution	equation (1.2)
turns out to be intermediate between
the heat equation   and the wave equation.
In general we refer to (1.2) as the fractional diffusion-wave equation,
and its solutions  can be interpreted as 
fractional diffusive waves, see Mainardi (1995).
\vsp
We point out that the viscoelastic models based on (2.14) or (2.15)
with $0<\nu <1$
and henceforth governed by  an evolution equation of fractional
order in time, see (1.2)  with $1/2<\beta  <1\,, $
are of great interest in material sciences and seismology.
In fact, as shown by Caputo and Mainardi (1971),
these models
exhibit an internal friction 
independent on frequency according to  the law
$$ Q^{-1} = {\rm tan}\, \l(\nu \,\pi\over 2\r) \,
\Longleftrightarrow \, \nu  = {2\over \pi}\, {\rm arctan}\,
   \l ( Q^{-1} \r) \,. \eqno(2.16)$$
The independence of the $Q$ from the frequency is in fact
experimentally verified  in pulse propagation phenomena
for many materials including  those of seismological interest.
From (2.16) we note that the $Q$ is also independent
on the material constants $\rho $ and $D\, $
which, however, play a role in the phenomenon of wave dispersion.
\vsp
The limiting cases of absence of energy dissipation (the elastic energy
is fully stored) and  of absence of energy storage (the elastic
energy is fully dissipated) are recovered from (2.16)
for $\nu =0\, $ (perfectly elastic solid)
and $\nu =1\,$ (perfectly viscous fluid), respectively.
\vsp
To obtain   values of seismological interest for the dissipation
 ($  Q \approx 1000$)	we need to choose the parameter $\nu $
sufficiently  close to zero, which corresponds
to a {\it nearly elastic} material; from (2.16) we obtain the
approximate relations between $\nu $ and $Q\,, $ namely
$$  \nu  \approx \l( {2\over \pi\, Q}\r)
     \approx 0.64 \, Q^{-1}  \,
  \Longleftrightarrow \,
      Q^{-1}  \approx {\pi\over 2}\, \nu
     \approx 1.57 \, \nu
\,. \eqno(2.17)$$
\vsp
As a matter of fact the evolution equation (1.2) turns out to be a
linear Volterra integro-differential equation of convolution
type with a weakly singular kernel of Abel type. Equations of
this kind have been treated,
both with and without reference to the fractional calculus,
by a number of authors	including
Caputo (1969, 1976, 1996b),
Meshkov and Rossikhin (1970),
Pipkin (1972-1986),
Buchen and Mainardi (1975),
Kreis and Pipkin (1986), Nigmatullin (1986),
Schneider and Wyss (1989),
Giona and Roman (1992), Metzler {\it et al.} (1994) and
Mainardi (1994, 1995, 1996a, 1996b).
For  recent reviews on related topics we refer to
Rossikhin and Shitikova (1997) and Mainardi (1997).


\section{The  Reciprocity Relation and the Auxiliary Functions}
\vs\vsp
The two basic problems for our fractional wave equation  (1.2)
concern, for $t \ge0$, the infinite interval $-\infty <x < +\infty$
and the semi-infinite interval $ x \ge 0\,, $ respectively;
the former is an initial - value problem, referred to as
the {\it Cauchy} problem, the latter is an initial boundary - value
problem, referred to as the {\it Signalling} problem.
\vsp
Extending
the classical analysis to our fractional equation (1.2), and
denoting by $g(x)$ and $h(t)$ two given, sufficiently well-behaving
functions,    the basic problems are thus formulated as following,
\vsh\pni
$a$) {\it Cauchy} problem,
$$   w(x,0^+)=g(x) \,, \q -\infty <x < +\infty\,; \qq
     w(\mp \infty,t) = 0\,,\q \, t>0\,;  \eqno(3.1a)
$$
\pni
$b$) {\it Signalling} problem,
$$  w(x, 0^+) =0 \,, \q  x>0\,;\qq
    w(0^+,t ) =h(t) \,, \q w(+\infty,t) =0 \,, \q   t >0 \,. \eqno(3.1b)
$$
\vsp
If $1/2 <\beta < 1\,, $ we must add in (3.1a) and (3.1b)
the initial values of the first time derivative of the field variable,
$w_t(x,0^+)\,,$ since in this case
(1.2) contains	a  time   derivative
of the second order.
To ensure the continuous dependence  of our solution
with respect to the parameter $\beta  $
also in the transition from $\beta  =(1/2)^-$ to  $\beta =(1/2)^+\,,$
we agree  to  assume $w_t(x,0^+) = 0\,. $
\vsp
In view of our analysis we find it convenient from now on to
add the parameter  $\beta $ to the
independent space-time variables $x\,,\,t$ in the solutions,
writing $w = w(x,t;\beta )\,. $
\vsp
For the {\it Cauchy} and {\it Signalling}  problems we
introduce
the so-called Green functions $\Gc (x,t;\beta )$ and $\Gs(x,t;\beta )$,
which represent the respective fundamental solutions,
obtained when $g(x) = \delta (x)$ and $h(t) = \delta (t)\,. $
As a consequence, the solutions of the two basic problems
are obtained by a space or time convolution  according to
$$ w(x,t;\beta)
= \int_{-\infty}^{+\infty} \Gc(x-\xi ,t;\beta	) \, g(\xi)
\, d\xi     \,,  \eqno(3.2a)$$
$$ w(x,t;\beta ) = \int_{0^-}^{t^+} \Gs(x,t-\tau;\beta  ) \, h(\tau ) \,
d\tau	  \,.  \eqno(3.2b)$$
It should be noted that in (3.2a) 
$\Gc(x ,t ;\beta) =   \Gc(|x|,t ;\beta)$ since the Green function of the Cauchy problem turns
out to be an even function of $x$.
According  to a usual convention,
in (3.2b) the limits of integration are extended to take into account
for the possibility of impulse functions centred at the extremes.
\vsp
For the  standard diffusion equation ($\beta  =1/2$)
it is well known that
$$   
\Gc(x,t;1/2) := \Gc^d (x,t)
 = {1\over 2\sqrt{\pi D }}\,t^{-1/2}\, \e^{-\ds \arg}\,,
 \eqno(3.3a) $$
$$\Gs(x,t;1/2) := \Gs^d (x,t)
 = {x\over 2\sqrt{\pi D }}\, t^{-3/2} \, \e^{-\ds \arg}\,.
 \eqno(3.3b) $$
In the limiting case $\beta =1$ we recover  the standard wave equation,
for which, putting $c = \sqrt{D }\,, $
$$   
\Gc(x,t;1) := \Gc^w (x,t)
 ={1\over 2} \l[  \delta (x-ct) + \delta (x+ct)\r] \,,
 \eqno(3.4a) $$
$$\Gs(x,t;1) := \Gs^w (x,t)
 = \delta (t-x/c)  \,. \eqno(3.4b)$$
\vsp
In the general case $0<\beta <1$ the two Green functions will be
determined by using the technique of the Laplace transform. This technique
  allows us  to obtain	the transformed
functions $\Gcs(x,s;\beta )$, $\Gss (x,s;\beta )$,
by solving ordinary differential equations of the 2-nd order in $x$
and then,
by inversion,  $\Gc(x,t;\beta )$ and $\Gs(x,t;\beta )$.
\vsp
 For the {\it Cauchy} problem (3.1a)
the application of the Laplace
transform to (1.2) with
$w(x,t)=\Gc(x,t;\beta)\, $
leads  to the non homogeneous differential equation
satisfied by the image of the Green function,
$\Gcs(x,s;\beta )\,, $
$$
 D\,{d^2 \Gcs  \over dx^2}-  s^{2\beta}  \,\Gcs=
   - \,\delta (x)\,s^{2\beta -1}\,, \q -\infty <x<+\infty
    \,.       \eqno(3.5)$$
Because of the singular term $\delta (x)$ we have to consider
the above equation separately in the two intervals $x<0$ and $x >0$,
imposing the   boundary conditions  at $x=\mp \infty\,, $
$\Gc({\mp \infty, t;\beta}) =0\,, $
and the necessary matching conditions
at $x= 0^{\pm}$.
We obtain
$$  \Gcs(x,s;\beta) =
    \rec{2\sqrt{D }\, s^{1-\beta}} \,
 \e^{\ds  -(|x|/\sqrt{D })s^{\beta}}\,, \q -\infty <x<+\infty
\,. \eqno(3.6)$$
 \vsp
 For the {\it Signalling} problem (3.1b)  the application of the Laplace
transform to (1.2) with
$w(x,t)=\Gs(x,t;\beta)\, $
 leads to   the  homogeneous differential equation
$$   D \,{d^2  \Gss \over dx^2}
   -s^{2\beta } \,\Gss	 = 0 \,, \q x\ge 0  \,.       \eqno(3.7)$$
Imposing the boundary conditions at $x=0\,, $
$\Gs(0^+,t;\beta)=h(t) =  \delta (t)\,, $
and at $x=+\infty\,, $
$\Gs(+\infty,t;\beta)=0\,, $
we obtain
$$  \Gss(x,s;\beta )  =
 \e^{\ds  -(x/\sqrt{D })s^{\beta }}\,, \q x\ge 0\,. \eqno(3.8)$$
From (3.6) and (3.8) we recognize
for the  original Green functions
 the following {\it reciprocity relation}
$$ 2\beta \, x\,  \Gc(x,t;\beta )  = t\, \Gs(x,t;\beta )
  \,,\q x > 0\,,\q t > 0 \,.\eqno(3.9) $$
This  relation	can be easily verified in the case of standard
diffusion ($\beta =1/2$),
where the explicit expressions
(3.4) of the Green   functions leads to the identity
    $$ x\, \Gc^d (x,t) = t\, \Gs^d (x,t) = {1\over 2\sqrt{\pi}}
   \, {x \over \sqrt{D \,t}}\,	\e^{-\ds\arg}=
F^d(r) =  {r\over 2}\,M^d(r)\,,
 \eqno(3.10)$$
where
  $r={x/(\sqrt{D}\, t^{1/2})} > 0\,$
is the well-known {\it similarity variable} and
$$ M^d(r)= \rec{\sqrt{\pi}} \,\e^{-\ds	r^2/4}\,.\eqno(3.11)$$
We   refer to $F^d(r)$ and $M^d(r)$ as
to the {\it auxiliary functions}
for the diffusion equation because each of them
provides the fundamental solution through (3.10).
We note that $M^d(r)$ satisfies the normalization condition
$ \int_0^{\infty} \! M^d(r)\, dr =1$.
\vsp
Applying  in the reciprocity relation (3.9)
 the complex inversion formula for the transformed Green functions
(3.6) and (3.8), and
changing the  integration variable in
 $  \sigma =s\,t\,,$
we obtain
$$ 2\beta \, x\,  \Gc(x,t;\beta )  = t\, \Gs(x,t;\beta )
    = F(r;\beta) = \beta r\, M(r;\beta )
  \,.\eqno(3.12) $$
where
$$r={x/(\sqrt{D}\, t^{\beta })} >0\,\eqno(3.13)$$
is the {\it similarity variable} and
$$
 F(r;\beta ) :=
 {1\over 2\pi i}\,\int_{Br}   \!\!
 \e^{\ds  \sigma -r\sigma ^\beta} \,
   {d\sigma} \,, \q
 M(r;\beta ) :=
 {1\over 2\pi i}\,\int_{Br}   \!\!
 \e^{\ds  \sigma -r\sigma ^\beta} \,
   {d\sigma \over \sigma^{1-\beta}} \,
    \eqno(3.14)$$
are the {\it auxiliary functions}.
In (3.14)  $Br$ denotes the Bromwich path
and $  r >0\,,$ $ 0<\beta <1\,.$
\newpage
 \vsp
The above  definitions of $F(r;\beta )$ and $M(r;\beta)$
by the {Bromwich representation} can be analytically continued
from $r>0$ to any $z \in \CC$,
by  deforming the Bromwich  path $Br$ into the Hankel path   $Ha\,, $
a contour that begins at $\sigma  =-\infty - ia$ ($a>0$), encircles the
branch cut that lies along the negative real axis, and ends up at
$\sigma = - \infty + ib$ ($b>0$).
\vsp
The integral and series representations
of $F(z;\beta )$ and  $M(z;\beta )$, valid on all of $\CC\,,$
with $0<\beta <1$
turn out to be
$$ 
\begin{array}{lll}
F(z;\beta ) & = 
{\ds {1\over 2\pi i}\,\int_{Ha}	\!\!
 \e^{\ds  \, \sigma -z\sigma ^\beta} \,
   d\sigma} \\  
 &= {\ds \sum_{n=1}^{\infty}{(-z)^n\over n!\, \Gamma(-\beta n)}}\\
  & = {\ds -{1\over \pi}\,\sum_{n=1}^{\infty}{(-z)^n\over n!}\,
    \Gamma(\beta n +1 )\, \sin(\pi \beta  n)}
 \end{array}          
   \eqno(3.15)$$
$$ 
\begin{array}{lll}
M(z;\beta ) & =  
 {\ds {1\over 2\pi i}\,\int_{Ha}   \!\!
 \e^{\ds  \,\sigma -z\sigma ^\beta} \,
   {d\sigma\over \sigma ^{1-\beta }}} \\
 & = {\ds \sum_{n=0}^{\infty} 
 {(-z)^n \over n!\, \Gamma[-\beta n + (1-\beta )]}}\\
 & =   {\ds \rec{\pi}\, \sum_{n=1}^{\infty}\,{(-z)^{n-1} \over (n-1)!}\,
  \Gamma(\beta n)  \,\sin (\pi\beta n)}	  
  \end{array}	
  \eqno(3.16)$$
\vsp
In the theory of special functions, see Erd\'elyi (1955),
we find an entire function, referred to as the
{\it Wright function},	which reads (in our notation)
  $$ W_{\lambda ,\mu }(z ) :=  {1\over 2\pi i}\,\int_{Ha}   \!\!
 \e^{\, \ds \sigma +z\sigma ^{-\lambda }} \,
   {d\sigma \over \sigma^{\mu}}
:=   \sum_{n=0}^{\infty}{z^n\over n!\, \Gamma(\lambda  n + \mu )}
   \,,	\q z \in \CC \,,
\eqno(3.17)$$
where $\lambda >-1$ and $\mu >0\,. $
From a comparison among (3.15-16) and (3.17) we recognize that
the auxiliary functions are  related to the Wright function according
to
$$ F(z;\beta ) =   W_{-\beta, 0}(-z) = \beta \, z \, M(z;\beta ) \,, \qq
   M(z;\beta ) =  W_{-\beta, 1-\beta}(-z)\,.
   \eqno(3.18)$$
Although convergent in all of $\CC$, the series representations
in (3.15-16) can be used to provide a numerical evaluation of our auxiliary
functions only for relatively small  values of $r\,, $ so that
asymptotic evaluations as $r \to +\infty$ are required.
Choosing as a variable $r/\beta $ rather than $r\,,$ the computation
by the saddle-point method for the $M$ function is
easier and yields, see Mainardi and Tomirotti (1995),
$$ M(r/\beta ;\beta ) \sim
   {r^{\ds{(\beta -1/2)/(1-\beta)}}
     \over \sqrt{2\pi\,(1-\beta)}}
  \,
   \exp  \l[- {1-\beta \over \beta }
       \,r^{\ds {1/(1-\beta)}}\r]\,,
 \;  r\to +\infty\,. \eqno(3.19)$$
We note that the saddle-point method
 for $\beta =1/2$  provides the
exact result (3.11), \ie
 $M(r;1/2) = M^d(r)= {\rm exp} (-r^2/4)/\sqrt{\pi}\,, $
but breaks down for $\beta \to 1^-.$
The case $\beta =1\,, $  for which
(1.2) reduces to the standard wave equation, is of course
a singular limit also for the series representation
since $M(r;1)= \delta (r-1)$. 
\vsp
The exponential decay for $r \to +\infty$ ensures that
all the moments of $M(r;\beta )$  in $\RR^+$ are finite;
in particular, see Mainardi (1997), we obtain
  $$ \int_0^{+\infty} \!\!\! r^{\, n}\, M(r;\beta )\, dr
   = {\Gamma(n+1)\over \Gamma(\beta n+1)}\,,\q n= 0\,,1\,, \, 2\, \dots
\eqno(3.20)$$
In Fig. 1 we exhibit plots of the auxiliary function
$M(r;\beta)$ in $\,0\le r\le 4\,$ for some rational
values of $\beta\,.$ 
The plots  are obtained by means of a numerical matching between
the series  and the saddle-point  representations.
As a matter of fact it turns out that the function $M(r;\beta )$ is
decreasing for $0<\beta < 1/2$, while
for $1/2<\beta <1$ it first increases and then decreases exhibiting
the maximum value $M_0(\beta )$ at a certain point $r_0(\beta)$;
as $\beta \to 1^-\,,\, M_0(\beta ) \to +\infty\,$ and
$r_0(\beta)\to 1\,. $
$\null$
 \vskip -0.75truecm
\noindent
\begin{figure}[h!]
\begin{center}
 \includegraphics[width=.90\textwidth]{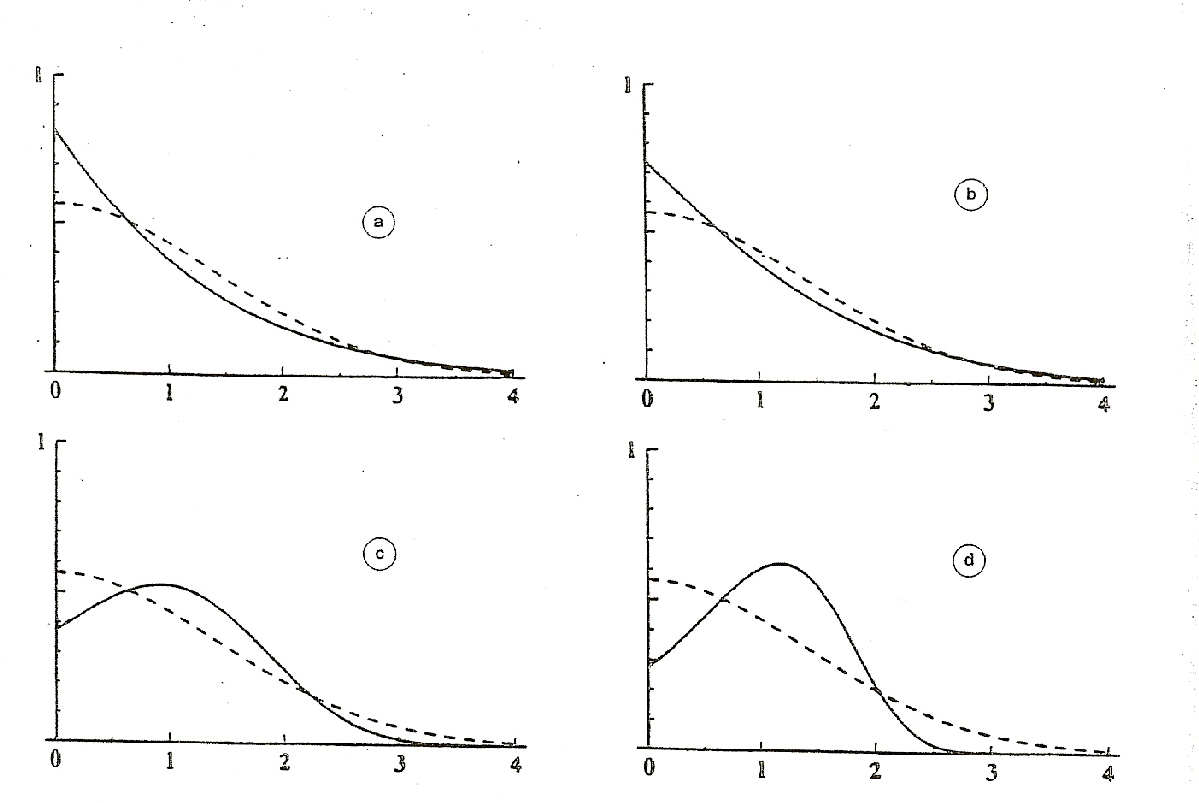}
\end{center}
\vskip -0.5truecm
 \caption{Comparison of $M(r;\beta )$ (continuous line)
with $M(r;1/2)$ (dashed line) in $0 \le r\le 4\,, $ for various value of
$\beta \,;$ (a) 1/4, (b) 1/3, (c) (2/3), (d) 3/4.
 \label{fig:1}}
 \end{figure}
$\null$
\vskip -2.5truecm
\section{The Evolution of the Seismic Pulse from the Fundamental
Solutions}
It is known that in theoretical seismology
the delta-Dirac function is of great relevance in simulating
the pulse generated by an ideal seismic source,
concentrated in space ($\delta (x)$) or in time
($\delta (t)$).
Consequently,
the fundamental solutions of the {\it Cauchy} and {\it Signalling}
problems are those of greater interest because they
provide us with information on the possible evolution
of the seismic pulses  during their propagation from the seismic
source.
Accounting of the reciprocity relation (3.12)
and  the similarity variable (3.13),
$r =  x/(\sqrt{D}\, t^\beta )\,, $
the two fundamental solutions can be written, for $x>0$
and $t>0\,, $
as
$$ \Gc(x,t;\beta ) = {1\over 2\beta x}\, F(r;\beta)
   =	{1\over2 \sqrt{D}\, t^\beta}\, M(r;\beta)
 \,,  \eqno(4.1a)$$
 $$\Gs(x,t;\beta ) = {1\over t}\, F(r;\beta)
 =  {\beta x\over \sqrt{D} \,t^{1+\beta}}\, M(r;\beta)\,.
  \eqno(4.1b)$$
\vfill\eject
The above equations mean that for the fundamental solution of the
 {\it Cauchy} [{\it Signalling}] problem the time [spatial] shape
is the same at each position [instant], the only changes
being due to space [time] - dependent changes of width and amplitude.
The maximum amplitude in time [space] varies precisely as
$1/x$ [$1/t$].
The two fundamental solutions  exhibit scaling properties
that  make easier their plots
versus distance  (at fixed instant) and versus time (at fixed position).
In fact,
 using the well-known scaling properties of the Laplace transform
 in (3.6) and (3.8),
  we easily prove, for any  $p\,,\,q >0\,,$ that
$$
 \Gc(px,qt;\beta ) = {1\over q^\beta}\, \Gc (px/q^\beta ,t;\beta)\,,
\eqno(4.2a) $$
$$\Gs(px,qt;\beta ) = {1\over q}\,  \Gs (px/q^\beta ,t;\beta)\,,
  \eqno(4.2b)$$
and, consequently, in plotting we can choose suitable values for the
fixed variable.
\begin{figure}[h!]
\begin{center}
 \includegraphics[width=.90\textwidth]{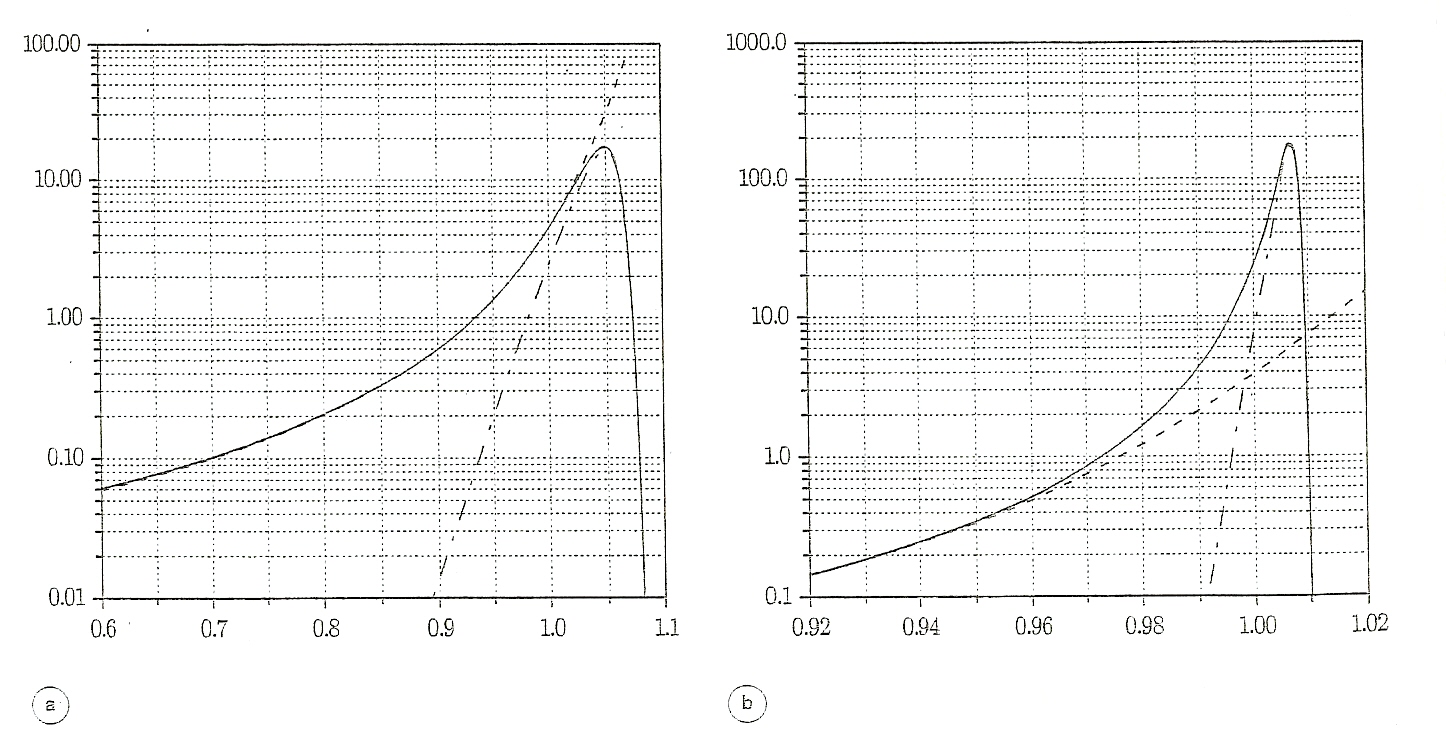}
\end{center}
\vskip -0.5truecm
 \caption{Comparison of the representations of $M(r;\beta )$
with $\beta =1-\epsilon$ around the maximum $r \approx 1\,, $
in the cases (a) $\epsilon =0.01\,,$ (b) $\epsilon =0.001\,,$
 obtained by Pipkin's method (continuous line), 100 terms-series (dashed line)  and
and saddle-point method (dashed-dotted line).
 \label{fig:2}}
 \end{figure}
\vsp
In order to inspect the evolution of the initial
pulse for seismological purposes,
we need to plot $ \Gc(x,t;\beta )$ versus $x$
and  $ \Gs(x,t;\beta )$ versus $t$  as $\beta $
is sufficiently close to $1$  ({\it nearly elastic} cases)
to ensure a sufficiently low value for the constant internal friction
$Q^{-1}\,. $  From (2.13)-(2.14) and (2.16)-(2.17)  we need to consider
$\beta =1-\epsilon $ with
$\epsilon  =\nu /2$ of the order of $0.001$ to	$0.01\,. $
In the evaluation of the auxiliary functions in the {\it nearly elastic}
cases, we note that the matching between the series and
saddle point representations is no longer achieved since the saddle
point turns out to be wide and the consequent approximation
becomes poor. In these cases we
need to adopt the ingenious variant of the saddle-point method introduced
by  Pipkin (1972-1986), see also Kreiss and Pipkin (1986), which
allows us to  see some structure in the peak while it tends
to the Dirac delta function. With Pipkin's method
we get the desired matching with the series representation
just in the region around the maximum $r \approx 1\,, $
as shown in Fig. 2a,b, where
we exhibit the significant plots of the auxiliary function
$M(r;\beta )$ with $\beta =1-\epsilon $ for $\epsilon =0.01\, $
and $\epsilon =0.001\,. $ 
\newpage
$\null$
\begin{figure}[h!]
\begin{center}
 \includegraphics[width=.90\textwidth]{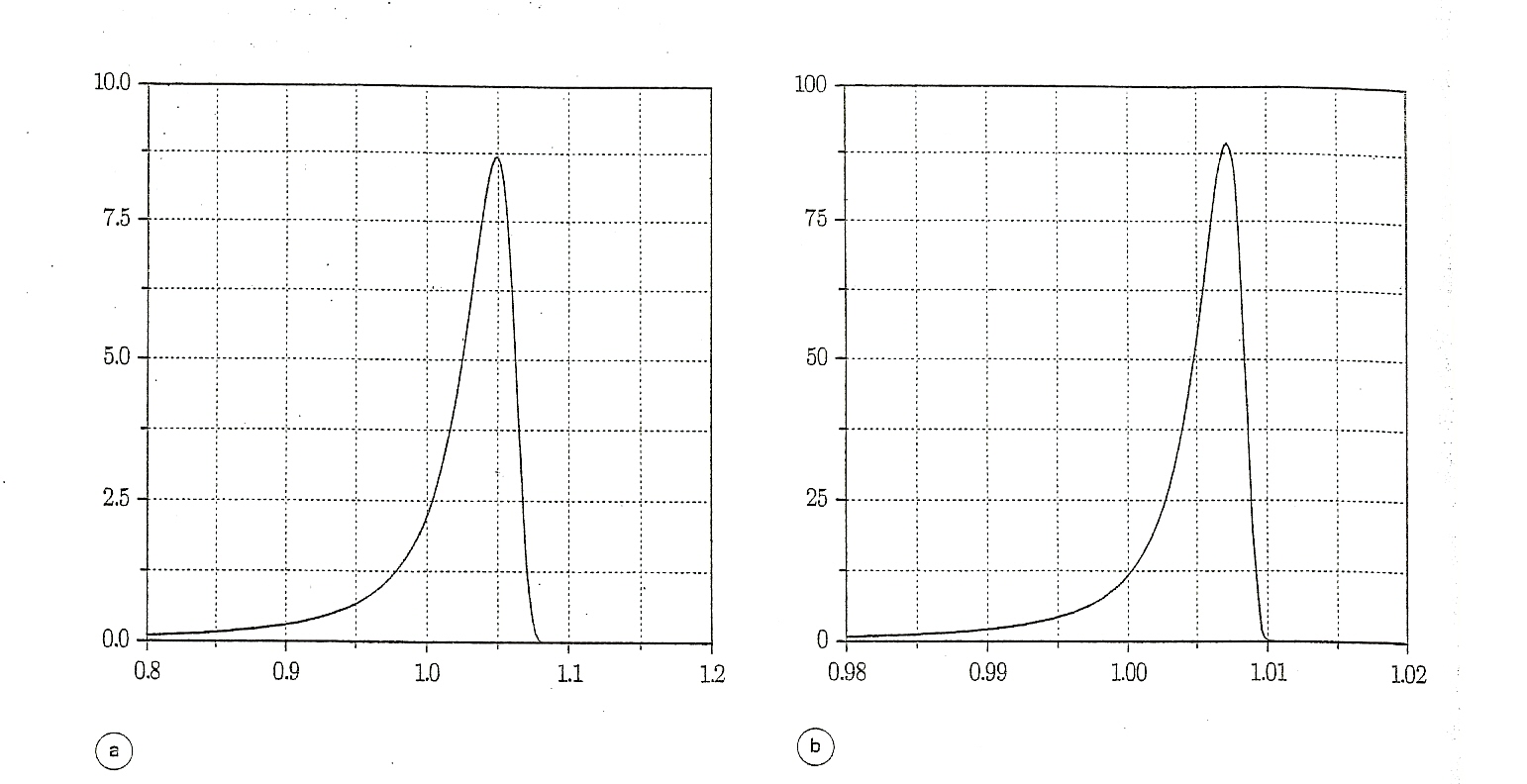}
\end{center}
\vskip -0.5truecm
 \caption{Plots of the fundamental solution $\Gc(x,t;\beta )$
versus $x$   at fixed $t=D=1\,, $
with $\beta =1-\epsilon$
in the cases (a) $\epsilon =0.01\,,$ (b) $\epsilon =0.001\,.$
 \label{fig:3}}
 \end{figure}
$\null$
\begin{figure}[h!]
\begin{center}
 \includegraphics[width=.90\textwidth]{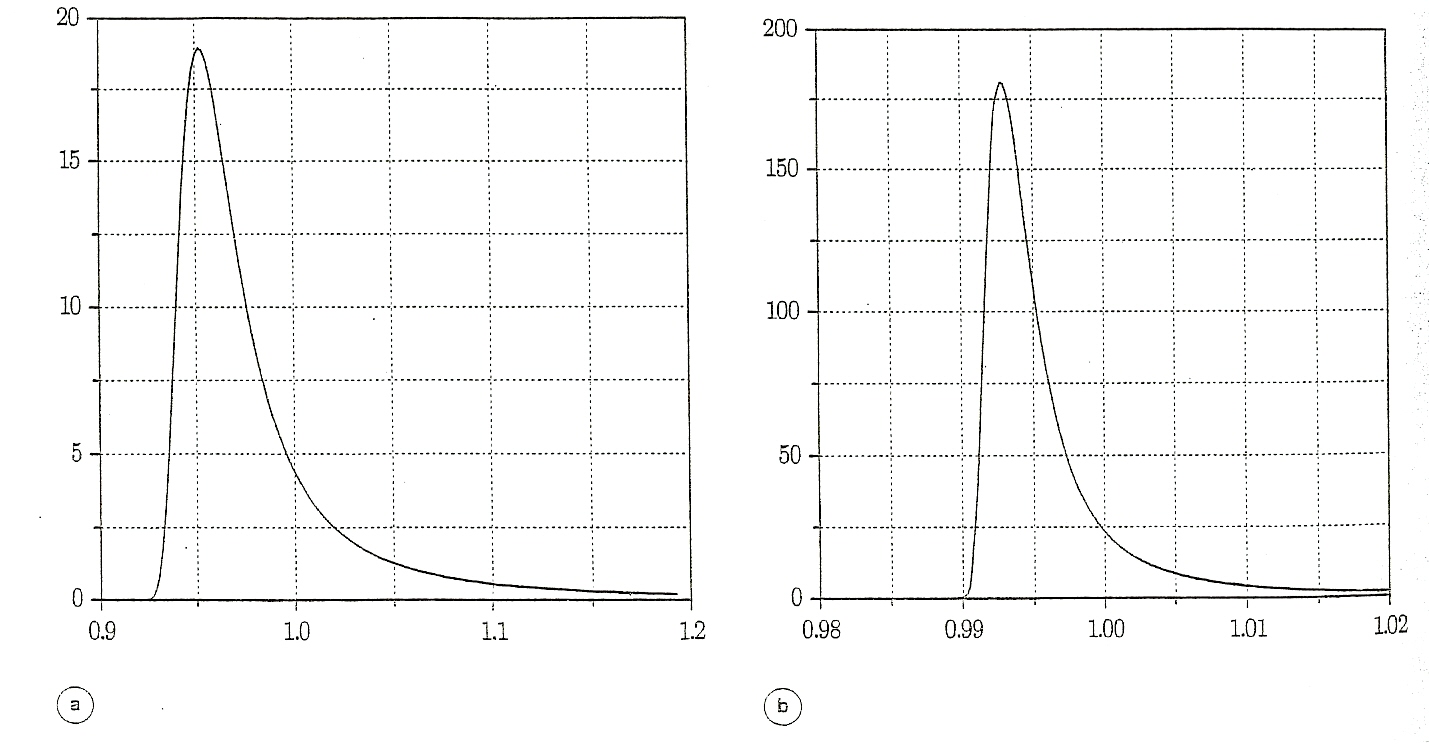}
\end{center}
\vskip -0.5truecm
 \caption{Plots of the fundamental solution $\Gs(x,t;\beta )$
versus $t$   at fixed $x=D=1\,, $
with $\beta =1-\epsilon$
in the cases (a) $\epsilon =0.01\,,$ (b) $\epsilon =0.001\,.$
\label{fig:4}}
\end{figure}
\vsp
Once obtained the auxiliary function $M(r;\beta)$
in the nearly elastic  cases,
we easily get the corresponding plots of the fundamental
solutions of the Cauchy and Signalling problems
by using (4.1a-b), see Figs. 3a,b and 4a,b.
\vsp
We also note  the  exponential decay of $\Gc(x,t;\beta)$
as $x \to +\infty$ (at fixed $t$) and the algebraic decay of
 $\Gs(x,t;\beta )$
as $t \to +\infty$ (at fixed $x$), for $0<\beta <1\,. $
In fact, using (4.1a,b) with (3.16) and (3.19),
we  get
$$  \Gc(x,t;\beta ) \sim a(t)\, x^{(\beta -1/2)/(1-\beta)}
   \, {\rm exp}\, \l[-b(t) x^{1/(1-\beta)}\r]\,, \q x\to \infty\,,
\eqno(4.3a)$$ $$
 \Gs(x,t;\beta ) \sim c(x)\, t^{-(1+\beta)} \,, \q
    t\to \infty\,, \eqno(4.3b)$$
where  $a(t)\,, \, b(t)$ and $c(x)$ are positive functions.

\section{ The Fundamental Solutions as  Probability  Density Functions}
\def\P{{\cal{P}}}
It is well known that the fundamental solution	of the standard diffusion
equation for the Cauchy problem is related with the
{\it Gauss} or {\it normal}
probability law, bilateral in space.
In fact, recalling (3.3a), we have
$$  \Gc^d (x,t) =
{1\over 2\sqrt{\pi\,D\,t}} \,	\e^{-\ds\arg} =
   p_G(x; \sigma ) \,, \eqno(5.1)$$
where
$$ p_G(x; \sigma ) :=
 {1\over \sqrt{2\pi}\,\sigma }\, \e ^{\,\ds -x^2/(2 \sigma ^2)}\,, \;
  \sigma^2 =2\, D\, t\,,
 \eqno(5.2) $$
denotes the well-known {\it Gauss} probability density function
($pdf$) with variance $\sigma ^2\,. $
The associated cumulative distribution function ($cdf$) is known to be
$$ 
\begin{array}{ll}
\P_G(x;\sigma ) & := {\ds \int_{-\infty}^x \!\!\!\! p_G(x;\sigma )\, dx} \\
  & = {\ds {\rec 2} \l[ 1 + {\rm erf} \l( {x\over \sqrt{2}\, \sigma }\r)\r]}
  = {\ds {\rec 2} \l[ 1 + {\rm erf} \l( {x\over 2\,\sqrt{D\,t}}\r)\r]}\,.
  \end{array}
  \eqno(5.3)$$
The  moments of   even order of the {\it Gauss} $pdf$
 turn out to be with $n \in \NN $,
$$ \int_{-\infty}^{+\infty} \!\!\!\!x^{2n}\,
  p_G (x;\sigma ) \,dx = {(2n)!\over 2^n \,n!}\, \sigma ^{2n}
  = (2n-1)!! \, \sigma ^{2n} = (2n-1)!! \, (2Dt)^n
   \,.
\eqno(5.4)$$
\vsp
If we consider	the fundamental solution  of the standard diffusion
equation but for the Signalling problem, we note that it is
related to the {\it L\'evy} probability law, unilateral in
time (a property not so well-known as that for the Cauchy problem!).
In fact, recalling (3.3b), we have
$$
 \Gs^d (x,t) =
 {x\over 2\sqrt{\pi\, D}\,t^{3/2}}
     \e^{-\ds\arg}=
   p_L(t; \mu )\,,   \eqno(5.5)$$
where
$$  p_L(t ; \mu ) =
   {\sqrt{\mu }\over \sqrt{2\pi}\, t^{3/2}}\, \e ^{\,\ds -\mu /(2t)}\,,
\;	\mu =  {x^2\over 2\,D}\,,
 \eqno(5.6)  $$
denotes the   {\it L\'evy} $pdf$ , see Feller (1971),
with $cdf$
$$ {\P}_L(t;\mu ) := \int_0^t \!\! p_L(t ;\mu  )\, dt  =
    {\rm erfc}\, \l( \sqrt{{\mu \over 2t}}\r)  =
 {\rm erfc} \,\l( {x\over 2\,\sqrt{D\,t}}\r)
      \,.   \eqno(5.7)$$
The {\it L\'evy} $pdf$ has all moments  of integer order infinite,
since it decays at infinity as $t^{-3/2}\,. $ However,
we note that
the moments of real order $\delta $ are finite only if
$ 0\le \delta  <1/2\,. $  In particular, for this $pdf$  the
mean (expectation) is infinite, but the {\it m\'ediane} is finite.
In fact, from ${\P}_L(t_{med};\mu)=1/2\,, $ it turns out that
$t_{med} \approx  2 \mu \,, $ since the complementary
error function gets the value 1/2 as its argument is
approximatively 1/2 (a better evaluation of the argument is 1/2.1).
\vsp
The {\it Gauss} and {\it L\'evy} laws 
are special cases of the important class of $\alpha$ - {\it stable}
probability distributions, or {\it stable} distributions with index
of stability  (or characteristic exponent)
$\alpha =2$ and $\alpha =1/2\,, $ respectively. Another special
case  is provided by the Cauchy law with $pdf$
$\, p_C(x;\lambda  )= \lambda /[\pi (x^2+\lambda ^2)]\,$
and $\alpha =1\,. $
\vsp
 The name stable has been assigned
to these distributions	because of
the following property:
if two independent real random variables
with the same shape or {\it type} of distribution are combined linearly and
the distribution of the resulting random variable has  the same shape,
the common distribution (or its type, more precisely) is said to be
{\it stable}.
More precisely,
if $Y_1$ and $Y_2$  are random variables
having such distribution, then $Y$ defined by the linear combination
$c\,Y = c_1\,Y_1 +c_2 \, Y_2$ has a similar distribution with the same
index $\alpha $ for any positive real values of the constants
$c\,,\,c_1$ and $c_2$ with
$c^\alpha = c_1^\alpha +c_2^\alpha \,. $
As a matter of fact only the range $0<\alpha \le 2$ is allowed for
the index of stability. The case $\alpha =2$ is noteworthy
since it corresponds to the {\it normal} distribution,
which is  the only stable distribution which has finite
variance, indeed finite moments of any order. In the cases $0<\alpha <2$
the corresponding $pdf$  $p_\alpha (y)$ have inverse power tails, \ie
$\int _{|y|>\lambda } p_\alpha (y)\,dy = O(\lambda ^{-\alpha })\,$
and therefore their    absolute moments
of order $\delta $ are finite if $0\le \delta <\alpha $ and
infinite if $\delta \ge \alpha \,. $
\vsp
The inspiration  for systematic research on stable distributions,
originated with Paul L\'evy,
was the desire to generalize the celebrated {\it Central Limit Theorem}
($CLT$).
\vsp
The restrictive condition of stability enabled some authors to derive
the general form for the characteristic function ($cf$, the
Fourier transform of the  $pdf$) of
a stable distribution, see Feller (1971).
A stable $cf$
is also {\it infinitely divisible}, \ie for every positive integer
$n$ it can be expressed as the $n$th power of some $cf$.
Equivalently we can say that for every positive integer $n$
a stable $pdf$ can be expressed as the	$n$-fold convolution of some
$pdf\,. $
All stable  $pdf$ are $unimodal$ and indeed {\it bell-shaped}, \ie
their $n$-th derivative has exactly $n$ zeros,
 \vsp
Using standardized random variables, the $\alpha $-{\it stable}
distributions turn out to depend on an additional parameter
 $\gamma \,, $
said the {\it skewness}  parameter.
Denoting a stable $pdf$
by $p_\alpha (y;\theta )\,, $
we note the   property
$ p_\alpha (-y;-\theta ) = p_\alpha (y;\theta )\,. $
Consequently a stable $pdf$ with  $\theta =0\, $
is necessarily symmetrical.
As a matter of fact $|\theta | \le \alpha $ if $0<\alpha <1$
and  $|\theta| \le 2- \alpha \, $  if $1<\alpha <2\,.$
 Stable distributions with extremal values of $\theta $ are
called {\it extremal}.
\vsp
From the theory one recognizes
that the normal distribution is the only stable  $df$ independent
on $\theta $,
and that all the extremal stable distributions with $0<\alpha  <1$
are unilateral, \ie vanishing in $\RR^\pm$ if $\theta =\pm \alpha \,. $
In particular, the following representations by convergent power series
are valid for stable distributions with $0<\alpha <1$ (negative powers)
and $1<\alpha < 2$ (positive powers),  for $y>0\,, $
$$ p_\alpha (y;\theta )=
{1\over \pi\,y}\,  \sum_{n=1}^{\infty}
   (-y^{-\alpha})^n \, {\Gamma (n\alpha +1)\over n!}\,
  \sin \l[{ n\pi\over 2}(\theta -\alpha)\r]\,,
 \;  0<\alpha <1\,, \eqno(5.8)$$
$$p_\alpha (y;\theta )=
{1\over \pi\,y}\,  \sum_{n=1}^{\infty}
   (-y)^{n} \, {\Gamma (n/\alpha +1)\over n!}\,
  \sin \l[{ n\pi\over 2\alpha }(\theta -\alpha)\r]\,,
 \;  1<\alpha <2\,.\eqno(5.9) $$
\vsp
From (5.8)-(5.9)  a relation    between stable $pdf$
with index $\alpha $ and $1/\alpha\,  $  can be derived.
Assuming $1/2<\alpha<1$ and $y>0\,, $  we obtain
$$ \rec{y^{\alpha +1}}\, p_{1/\alpha}(y^{-\alpha} ;\theta )
  =p_\alpha (y;\theta ^*)\,,  \;
  \theta ^*=\alpha(\theta +1)-1 \,. \eqno(5.10)$$
A quick check shows that  $\theta^*$ falls within the prescribed range,
$|\theta ^*|\le\alpha \,, $ provided that $|\theta |\le 2-1/\alpha \,. $
Furthermore, we can derive a relation between extremal stable $pdf$
and  our auxiliary functions 
of Wright type.
In fact, by comparing (5.8-9) with the series representations in
(3.15-16) and using (3.18),   we obtain
$$
p_\alpha (y;-\alpha )
 =  \rec{y}\,  F(y^{-\alpha };\alpha) =
 {\alpha \over y^{\alpha +1}}\,  M(y^{-\alpha };\alpha ) \,,\;
 0<\alpha <1\,, \eqno(5.11)$$
$$ p_\alpha (y;\alpha -2)
 = \rec{y}\,  F(y;1/\alpha) =
 {1 \over \alpha }\,  M(y;1/\alpha ) \,,\;
 1<\alpha <2\,.\eqno(5.12)$$
Consequently  we can interpret the fundamental	solutions (4.1a) and
(4.1b) in terms of stable $pdf$, so generalizing the arguments
for the standard diffusion  equation
based on (5.1-7).
\vsp
We easily recognize
that for $0<\beta <1$
the fundamental solution for the Signalling problem  provides
a unilateral extremal stable $pdf$ in (scaled) time with index of
stability $\alpha = \beta \,, $  which decays according to (4.3b)
with a power law.
In fact, from  (4.1b) and (5.11) we note that,
putting $y=r^{-1/\beta }= \tau \,, $
$$ (x/\sqrt{D})^{1/\beta}\, \Gs(x,t;\beta ) =
p_\beta (\tau;-\beta )\,,
\q \tau =t \,({\sqrt{D}/ x})^{1/\beta } >0   \,. \eqno(5.13)
$$
This property has been	noted also by Kreiss and Pipkin (1986)
based on (3.8) and  on Feller's   result,
$\, p_\alpha  (t; -\alpha  )  \div {\rm exp} (-s^\alpha) $  for
$0<\alpha <1\,. $
\vsp
As far as the Cauchy problem
is concerned, we note that the corresponding fundamental
solution provides  a bilateral symmetrical $pdf$ in (scaled) distance
with two 
branches, for $x>0$ and $x<0\,, $
obtained one from the other by reflection.
For large $|x|$ each branch exhibits an exponential decay according to
(4.3) and, only for $1/2\le \beta <1\,, $
it is the corresponding branch of an extremal  stable $pdf$
with index of stability $\alpha =1/\beta \,. $
In fact, from (4.1b) and (5.12) we note  that, putting
$y= |r| =\xi >0\,, $
$$ 2\beta \, \sqrt{D}\,t^\beta \,  \Gc(|x|,t;\beta )
 =  p_{1/\beta }(\xi, 1/\beta -2) \,,
\q \xi= |x|/(\sqrt{D}\, t^\beta) >0\,. \eqno(5.14) $$
This property had to the authors' knowledge not been noted:
it properly generalizes the Gaussian property of the $pdf$
found for   $\beta =1/2\, $ (standard diffusion).
Furthermore, using (3.20),  the moments (of even order)
of $\Gc(x,t;\beta)$ turn out to be
$$ \int_{-\infty}^{+\infty} \!\!\! x^{2n}\, \Gc(x,t;\beta )  \,dx =
   {\Gamma(2n+1)\over \Gamma(2\beta n+1)}\, (D t^{2\beta })^n\,,
\; n=0\,, \,1\,, \,2\,, \, \dots
\eqno(5.15)$$
We recognize that the variance is now proportional to $D t^{2\beta }\,, $
which implies a phenomenon of {\it fast} diffusion if $1/2<\beta <1\,. $

 \section*{Appendix: Essentials of Fractional Calculus}
Fractional calculus is the field of mathematical analysis which deals
with the investigation and applications of integrals and
derivatives of arbitrary order.
The term {\it fractional} is a misnomer, but it is
retained 
following the  prevailing use.
\vsp
According to the Riemann-Liouville approach  to fractional calculus,
the notion of fractional integral of order $\alpha$
($\alpha >0$)
 is a natural consequence
of the well known formula (usually attributed to Cauchy),
that reduces the calculation of the $n-$fold primitive of a function
$f(t)$ to a single integral of convolution type.
In our notation the Cauchy formula reads
$$
    J^n f(t) := f_{n}(t)=
 \rec{(n-1)!}\, \int_0^t \!\!  (t-\tau )^{n-1}\,f(\tau) \, d\tau\,,
    \q t > 0\,,\q n \in \NN   \,, \eqno(A.1) $$
where $\NN$ is the set of positive integers.
From this definition we note that $f_{n}(t)$
vanishes at $t=0$ with its
derivatives of order $1,2, \dots, n-1\,. $
For convention we require that	$f(t)$ and henceforth
$f_{n}(t)$ be a {\it causal} function, \ie identically
vanishing for $t<0\,. $
\vsp
In a natural way  one is  led
to extend the above formula
from positive integer values of the index to any positive real values
by using the Gamma function.
Indeed, noting that $(n-1)!= \Gamma(n)\,, $
and introducing the arbitrary {\it positive} real number
 $\alpha\,, $
one defines  the
 \ \underbar{{\it Fractional Integral of order} $\alpha >0 $} :
$$
J^\alpha \,f(t) :=
      \rec{\Gamma(\alpha )}\,
 \int_0^t (t-\tau )^{\alpha -1}\, f(\tau )\,d\tau \,,
   \q t > 0\,,\q \alpha  \in \RR^+
 \,,  \eqno(A.2) $$
where $\RR^+$ is the set of positive real numbers.
For complementation we define
$J^0 := I\, $ ({Identity operator)}, \ie we mean
$J^0\, f(t) = f(t)\,. $ Furthermore,
by $J^\alpha f(0^+)$ we mean the limit (if it exists)
of $J^\alpha f(t)$ for $t\to 0^+\,;$ this limit may be infinite.
\vsp
We note the {\it semigroup property}
$J^\alpha J^\beta = J^{\alpha +\beta}\,,
   \; \alpha\,,\;\beta	\ge 0\,,$
which implies the {\it commutative property}
$J^\beta  J^\alpha= J^\alpha J^\beta\,,$
and  the effect of our operators $J^\alpha$
on the power functions
$$
J^\alpha t^\gamma ={\Gamma (\gamma +1)\over \Gamma(\gamma +1+\alpha)}\,
		   t^{\gamma+\alpha}\,, \q \alpha \ge 0\,,
  \q \gamma >-1\,, \q t>0\,.
\eqno (A.3)
$$
These properties  are of course a natural generalization
of those known when the order is a positive integer.
\vsp
Introducing
the Laplace transform by the notation
$ {\cal{L}}\, \l\{  f(t) \r\}  := \int_0^\infty \!\!
   \e^{-st}\, f(t)\, dt = \widetilde f(s)\,, \; s \in \CC\,,$
and  using the sign $\div$ to denote a Laplace transform pair,
\ie
$ f(t) \div  \widetilde f(s) \,, $
we note the following rule for the   Laplace transform of
the fractional integral,
$$	   J^\alpha \,f(t) \div
     {\widetilde f(s)\over s^\alpha}\,,\q \alpha \ge 0\,,  \eqno(A.4)$$
which is the generalization
of the case with an $n$-fold repeated integral. 
\vsp
After the notion of fractional integral,
that of fractional derivative of order $\alpha$
($\alpha >0$)
becomes a natural requirement and one is attempted to
substitute $\alpha $ with $-\alpha $ in the above formulas.
However, this generalization  needs some care  in order to
guarantee the convergence of  the integrals   and
preserve the
well known properties of the ordinary derivative of integer
order.
\vsp
 Denoting by $D^n\,$ with $ n\in \NN\,, $
the operator of the derivative of order $n\,,$	we first note that
$ D^n \, J^n = I\,, \;	 J^n \, D^n \ne I\,,\q n\in \NN \,,
$
\ie $D^n$ is left-inverse (and not right-inverse) to
the corresponding integral operator $J^n\,. $
In fact we easily recognize from (A.1) that
$$  J^n \, D^n \, f(t) = f(t) - \sum_{k=0}^{n-1}
	f^{(k)}(0^+) \, {t^k\over k!}\,, \q t>0\,. \eqno(A.5)$$
As a consequence we expect that $D^\alpha $ is defined as left-inverse
to $J^\alpha $.  For this purpose, introducing the positive
integer $m$ such that $m-1 <\alpha \le m\,, $
one defines the
 \underbar{{\it Fractional Derivative of order} $\alpha >0 $}
as $\; D^\alpha \,f(t) := D^m \, J^{m-\alpha} \, f(t)\,,$
\ie
$$ \!\!\!
 D^\alpha \,f(t) :=
\cases{
  {\ds {d^m\over dt^m}}\l[
  {\ds \rec{\Gamma(m-\alpha)}\int_0^t
    {f(\tau )\over (t-\tau )^{\alpha +1-m}} \,d\tau}\r] ,
 & $ m-1 <\alpha < m,$ \cr\cr
     {\ds {d^m\over dt^m}} f(t)\,,
    & $ \alpha =m\,. $\cr\cr }
   \eqno(A.6) $$
Defining for complementation $D^0 = J^0 =I\,, $ then
we easily recognize that
$ D^\alpha \, J^\alpha = I \,,$  $\, \alpha \ge 0\,,$
and
$$ D^{\alpha}\, t^{\gamma}=
   {\Gamma(\gamma +1)\over\Gamma(\gamma +1-\alpha)}\,
     t^{\gamma-\alpha}\,,
 \q \alpha \ge 0\,,
  \q \gamma >-1\,, \q t>0\,.
\eqno (A.7)
$$
Of course, these properties are a natural generalization
of those known when the order is a positive integer.
\vsp
Note the remarkable fact that the fractional derivative $D^\alpha\, f$
is not	zero
for the constant function $f(t)\equiv 1$ if $\alpha \not \in {\NN}\,. $
In fact, (A.7) with $\gamma =0$ teaches us that
$$
D^\alpha 1 = {t^{-\alpha}\over \Gamma(1-\alpha)}\,,\q \alpha\ge 0\,,
\q t>0\,.  \eqno (A.8)
$$
This, of course, is $\equiv 0$ for $\alpha \in{\NN}$, due to the
poles of the gamma function in the points $0,-1,-2,\dots$.
We now observe that an alternative definition
of fractional derivative, originally introduced by Caputo (1967)
(1969)
in the late sixties and
adopted by Caputo and Mainardi (1971)
in the framework  of the theory of {\it Linear Viscoelasticity},
is
$\; D_*^\alpha	\, f(t) :=   J^{m-\alpha}\, D^{m} \, f(t) $
with $m-1 <\alpha \le m\,, \; m\in \NN\,,$ \ie
$$
 D_*^\alpha \,f(t) :=
\cases{
    {\ds \rec{\Gamma(m-\alpha)}}\,{\ds\int_0^t
 {\ds {f^{(m)}(\tau)\over (t-\tau )^{\alpha +1-m}}} \,d\tau} \,,
  & $\; m-1<\alpha <m\,, $\cr\cr
     {\ds {d^m\over dt^m}} f(t)\,,
    & $\; \alpha =m\,. $\cr\cr }
   \eqno(A.9) $$
This definition is of course more restrictive than (A.6), in that
requires the absolute integrability of the  derivative of order $m$.
Whenever we use the operator   $D_*^\alpha$   we (tacitly) assume that
     this condition is met.
 We  easily recognize that in general
$$  D^\alpha\, f(t) := D^{m} \, J^{m-\alpha} \, f(t)
 \ne J^{m-\alpha}\, D^{m} \, f(t):= D_*^\alpha \, f(t)\,,
 \eqno(A.10)
 $$
 unless   the function	$f(t)$ along with its first $m-1$ derivatives
 vanishes at $t=0^+$.
In fact, assuming that
the passage of the $m$-derivative under
the integral is legitimate, one     
recognizes that,  for $ m-1 <\alpha  < m \,$  and $t>0\,, $
$$
    D^\alpha \, f(t) =
  D_*^\alpha   \, f(t) +
  \sum_{k=0}^{m-1}   {t^{k-\alpha}\over\Gamma(k-\alpha +1)}
    \, f^{(k)}(0^+) \,, \eqno(A.11)    $$
 and therefore, recalling the fractional derivative of the power
functions (A.7),
$$
   D^\alpha \l( f(t) -
 \sum_{k=0}^{m-1} {t^k \over k!} \, f^{(k)} (0^+)\r)
     =	D_*^\alpha  \, f(t)  \,.\eqno(A.12)  $$
The alternative definition (A.9) for the
fractional derivative  thus incorporates the initial values
of the function and of its integer derivatives of lower order.
The subtraction of the Taylor polynomial of degree $m-1$ at $t=0^+$
from $f(t)$ means  a sort of
regularization	of the fractional derivative.
In particular, according to this definition,
the relevant property for which the fractional derivative
of a constant is still zero can be easily recognized,
 \ie
$$ D_*^\alpha  1 \equiv 0\,,\q	 \alpha >0\,.\eqno(A.13)$$
   \vsp
We now explore the most relevant differences between the two
fractional derivatives (A.6) and (A.9). We agree to
denote (A.9) as the {\it Caputo fractional derivative}
to distinguish it from the standard Riemann-Liouville fractional
derivative (A.6).
We observe, again by looking at (A.7), that
$D^\alpha t^{\alpha -1} \equiv 0\,, \; \alpha>0\,, \; t>0\,.$
From above  we thus recognize
the following statements about functions
which  for $t>0\, $   admit the same fractional derivative
of    order $\alpha \,, $
with $m-1 <\alpha \le m\,,$ $\; m \in \NN\,, $
$$    D^\alpha \, f(t) = D^\alpha  \, g(t)
   \,  \Longleftrightarrow  \,
  f(t) = g(t) + \sum_{j=1}^m c_j\, t^{\alpha-j} \,,
    \eqno(A.14) $$
$$    D_* ^\alpha \, f(t) = D_*^\alpha	\, g(t)
   \,  \Longleftrightarrow  \,
  f(t) = g(t) +  \sum_{j=1}^m c_j\, t^{m-j} \,.
    \eqno(A.15) $$
In these formulas the coefficients $c_j$ are arbitrary constants.
\vsp
For the two definitions we also note a difference
with respect to the {\it formal} $\,$ limit  as
 $\alpha \to {(m-1)}^+$. From (A.6) and (A.9) we obtain
respectively,
$$ \alpha \to (m-1)^{+}\,\Longrightarrow\,
 D^\alpha \,f(t) \to	D^m\,  J\, f(t) = D^{m-1}\, f(t)
   \,; \eqno(A.16) $$
$$\alpha \to {(m-1)}^{+} \,\Longrightarrow\,
 D_*^\alpha \, f(t) \to J\, D^m\, f(t) =
       D^{m-1}\, f(t) - f^{(m-1)} (0^+)\,. \eqno(A.17) $$
\vsp
We now consider the {\it Laplace transform} of the two fractional
derivatives.
For the standard fractional derivative $D^\alpha $
the Laplace transform,	assumed to exist,  requires the knowledge of the
(bounded) initial values of the fractional integral $J^{m-\alpha }$
and of its integer  derivatives of order $k =1,2, \dots, m-1\,. $
The corresponding rule reads, in our notation,
$$ D^\alpha \, f(t) \div
      s^\alpha\,  \widetilde f(s)
   -\sum_{k=0}^{m-1}  D^k\, J^{(m-\alpha)}\,f(0^+) \, s^{m -1-k}\,,
  \q m-1<\alpha \le m \,. \eqno(A.18)$$
\vsp
The {\it Caputo fractional derivative} appears more suitable to
be treated by the Laplace transform technique in that it requires
the knowledge of the (bounded)
initial values of the function
and of its integer  derivatives of order $k =1,2, \dots, m-1\,, $
in analogy with the case when $\alpha =m\,. $
In fact,
by using (A.4) and noting that
$$\! \!\!\! J^\alpha  \, D_*^\alpha \, f(t) =
    J^\alpha\, J^{m-\alpha }\, D^m \, f(t) =
     J^m\, D^m \, f(t) = f(t) -
  \sum_{k=0}^{m-1} {f^{(k)}(0^+)} {t^k \over k!}
  . \eqno(A.19)$$
we easily prove  the following rule for the Laplace transform,
$$ D_*^\alpha \, f(t) \div
      s^\alpha\,  \widetilde f(s)
   -\sum_{k=0}^{m-1}  f^{(k)}(0^+) \, s^{\alpha -1-k}\,,
  \q m-1<\alpha \le m \,. \eqno(A.20)$$
Indeed, the  result (A.20), first stated by Caputo (1969) by using the
Fubini-Tonelli theorem, appears  as the most "natural"
generalization of the corresponding result well known for $\alpha =m\,. $
\vsp
This appendix is  based on the review by Gorenflo and Mainardi
(1997). For more details on the classical treatment of	fractional
calculus
the reader is referred to Erd\'elyi (1954), Oldham and Spanier (1974),
Samko {\it et al.} (1987-1993) and  Miller and Ross (1993).
Gorenflo and Mainardi
have pointed out the major utility of the
Caputo fractional derivative
in the treatment of differential equations of fractional
order for {\it physical applications}.
In fact, in physical problems,	the initial conditions are usually
expressed in terms of a given number of bounded values assumed by the
field variable and its derivatives of integer order,
no matter if
the governing evolution equation may be a generic integro-differential
equation and therefore, in particular,	a  fractional differential
equation.
\section*{References}
\vsp
\rl Aki, K. and P.G. Richards (1980):
  {\it Quantitative Seismology} (Freeman, San Francisco),
  Vol. 1,  Ch. 5, pp. 167-185.
\vsp
\rl Ben-Menahem, A., and S.J. Singh (1981):
  {\it Seismic Waves and Sources} (Springer-Verlag, New York),
Ch. 10, pp. 840-944.
\vsp
\rl Buchen, P.W. and  F. Mainardi (1975):
    Asymptotic expansions for transient viscoelastic waves,
    {\it  J.  M\'ec.}  {\bf 14}, 597-608.
\vsp
\rl  Caputo, M. (1966) :
  Linear models of dissipation whose Q is almost frequency independent,
  {\it Annali di Geofisica}, {\bf 19}, 383-393.
\vsp
\rl Caputo, M. (1967) :
  Linear models of dissipation whose Q is almost frequency independent,
  Part II.,
  Geophys. J. R. Astr. Soc., {\bf 13}, 529-539.
\vsp
\rl Caputo, M.	(1969): {\it Elasticit\`a e Dissipazione}
  (Zanichelli Bologna). [in Italian]
\vsp
\rl Caputo, M. and F. Mainardi (1971):
Linear models of dissipation in anelastic solids,
 {\it Riv. Nuovo Cimento} (Ser II) {\bf 1}, 161-198.
\vsp
\rl  Caputo, M. (1976):
  Vibrations of an infinite plate  with a frequency
  independent $Q\,, $
 {\it J. Acoust. Soc. Am.}, {\bf 60}, 634-639.
\vsp
\rl  Caputo, M. (1979):
  A model for the fatigue in elastic materials with frequency
  independent $Q\,, $
 {\it J. Acoust. Soc. Am.}, {\bf 66}, 176-179.
\vsp
\rl  Caputo, M. (1981):
  Elastic radiation from a source in a medium with an almost frequency
  independent $Q\,,$
  {\it J. Phys.  Earth}, {\bf 29}, 487-497.
\vsp
\rl  Caputo, M. (1985) :
  Generalized rheology and geophysical consequences,
{\it Tectonophysics}, {\bf 116}, 163-172.
\vsp
\rl Caputo, M. (1996)a:
  Modern rheology and electric induction:
multivalued index of refraction, splitting of eigenvalues and fatigues,
  {\it Annali di Geofisica}, {\bf 39}, 941-966.
\vsp
\rl Caputo, M. (1996)b:
  The Green function of the diffusion in porous media with memory,
  {\it Rend. Fis. Acc. Lincei} (Ser. 9), {\bf 7}, 243-250.
\vsp
\rl Carcione, J.M., Kosloff, D. and R. Kosloff (1988):
   Wave propagation in a linear viscoelastic medium,
  {\it Geophys. J.}, {\bf95}, 597-611.
\vsp
\rl  Chin, R.C.Y. (1980):
    Wave propagation in viscoelastic media,
  in {\it Physics of the Earth's Interior},
  edited by A. Dziewonski and E. Boschi
  (North-Holland, Amsterdam), pp. 213-246.
 [E. Fermi Int. School, Course 78]
\vsp
\rl Christensen, R.M. (1982): {\it Theory of Viscoelasticity}
(Academic Press, New York).  [1-st ed. (1972)]
\vsp
\rl Erd\'elyi, A. {\it Editor} (1954):
  {\it Tables of Integral Transforms},	Bateman Project
  (McGraw-Hill, New York), Vol. 2, Ch. 13,  pp. 181-212.
\vsp
\rl  Erd\'elyi, A. {\it Editor} (1955):
  {\it Higher Transcendental Functions},
  Bateman Project (McGraw-Hill, New York), Vol. 3, Ch. 18, pp. 206-227.
\vsp
\rl Feller, W. (1971),
{\it An Introduction to Probability Theory and its Applications},
  (Wiley, New York), Vol. II, Ch. 6: pp. 169-176, Ch. 13: pp. 448-454.
[1-st ed. (1966)]
\vsp
\rl  Futterman, W.I. (1962):  Dispersive Body Waves,
 {\it J. Geophys. Res.}, {\bf 67},  5279-5291.
\vsp
\rl Gel'fand, I.M.  and G.E. Shilov (1964): {\it Generalized Functions},
 (Academic Press, New  York), Vol. I.
\vsp
\rl  Giona, M.	and  H.E. Roman (1992):
  Fractional diffusion equation for transport phenomena in random
  media,  {\it Physica A}, {\bf 185},  82-97.
\vsp
\rl Gordon, R.B. and C.W. Nelson (1966):
  Anelastic properties of the Earth,
 {\it Rev. Geophys.}, {\bf 4}, 457-474 (1966).
\vsp
\rl Gorenflo, R. and F. Mainardi (1997):
 Fractional calculus: integral and differential
 equations of fractional order, in
 {\it Fractals and Fractional Calculus in Continuum Mechanics}, edited by
 A. Carpinteri and F. Mainardi (Springer Verlag, Wien), 223-276.
\vsp
\rl  Graffi, D. (1982):
  Mathematical models and waves in linear viscoelasticity,
  in {\it Wave Propagation in Viscoelastic Media}, edited by F. Mainardi
  (Pitman, London), pp. 1-27.  [Res. Notes in Maths, Vol. 52]
 \vsp
\rl Hunter, S.C. (1960): Viscoelastic Waves, in
 {\it Progress in Solid Mechanics}, edited by
  I. Sneddon and R. Hill
  (North-Holland, Amsterdam),  Vol 1, pp. 3-60.
\vsp
\rl Jackson, D.D. and D.L. Anderson (1970):
  Physical mechanisms for seismic wave attenuation,
 {\it Rev. Geophys.}, {\bf 2}, 625-660 (1964).
\vsp
\rl Kanamori, H. and D.L. Anderson (1977):
 Importance of physical dispersion in surface wave and free oscillation
 problems,
 {\it Rev. Geophys.}, {\bf 15}, 105-112 (1977).
\vsp
\rl Kang, I.B. and G.A. McMechan (1993):
Effects of viscoelasticity on wave propagation in fault zones,
near-surfaces sediments and inclusions,
 {\it Bull. Seism. Soc. Am.}, {\bf 83}, 890-906.
\vsp
\rl Kjartansson, E. (1979):
 Constant-$Q$ wave propagation and attenuation,
 {\it J. Geophys. Res.}, {\bf 94}, 4737-4748.
\vsp
\rl Knopoff, L. (1964):
  $Q\,, $ {\it Rev. Geophys.}, {\bf 2}, 625-660.
\vsp
\rl Kolsky, H. (1956): The propagation of stress pulses in viscoelastic
 solids, {\it Phil. Mag.} (Ser 8), {\bf 2}, 693-710.
\vsp
\rl K\"ornig, M. and G. M\"uller (1989):
Rheological models and interpretation of postglacial uplift,
{\it Geophys. J. Int.}, {\bf 98}, 243-253.
\vsp
\rl Kreis, A. and A.C. Pipkin (1986):
  Viscoelastic pulse propagation and stable probability distributions,
  {\it Quart. Appl. Math.}, {\bf 44}, 353-360.
\vsp
\rl Mainardi, F. and  G. Turchetti (1975):
  Wave front expansion for transient viscoelastic waves,
 {\it Mech. Res. Comm.}  {\bf 2}, 107-112.
\vsp
\rl Mainardi,  F. (1994):
 On the initial value problem for the fractional
  diffusion-wave equation, in
  {\it Waves and Stability in Continuous Media}
  edited by S. Rionero and T. Ruggeri,
  (World Scientific, Singapore), pp. 246-251.
\vsp
\rl Mainardi,  F. (1995):
  Fractional diffusive waves  in viscoelastic solids
  in  {\it IUTAM Symposium - Nonlinear Waves in Solids},
 edited by J. L. Wegner  and F. R. Norwood (ASME/AMR, Fairfield NJ),
  pp. 93-97.
 [Abstract in {\it Appl. Mech. Rev.}, {\bf 46} (1993), 549]
\vsp
\rl Mainardi, F. and M. Tomirotti (1995): On a special function
  arising  in  the time fractional diffusion-wave equation,
  in {\it Transform Methods and Special Functions, Sofia 1994},
edited by P. Rusev, I. Dimovski and V. Kiryakova,
  (Science Culture Technology, Singapore), pp. 171-183.
\vsp
\rl Mainardi, F. (1996)a:
 Fractional relaxation-oscillation and fractional
  diffusion-wave phenomena,
  {\it Chaos, Solitons \& Fractals}, {\bf 7}, 1461-1477.
\vsp
\rl Mainardi, F. (1996)b:
  The fundamental solutions for the fractional diffusion-wave
   equation,
   {\it Applied Mathematics Letters}, {\bf 9}, No 6, 23-28.
\vsp
\rl    Mainardi, F. (1997):
 Fractional calculus;
  some basic problems in continuum and statistical mechanics,
  in {\it Fractals and Fractional Calculus in Continuum Mechanics},
edited by. A. Carpinteri and F. Mai\-nardi
 (Springer-Verlag, Wien), 291-348.
\vsp
\rl Meshkov, S.I. and Yu. A. Rossikhin (1970):
Sound wave propagation in a viscoelastic medium whose hereditary
properties are determined by weakly singular kernels, in
{\it Waves in Inelastic Media}, edited by Yu. N. Rabotnov
(Kishniev), pp. 162-172. [in Russian]
\vsp
\rl  Metzler, R.,  Gl\"ockle, W.G. and T.F. Nonnenmacher (1994):
   Fractional model equation for anomalous diffusion,
   {\it Physica A}, {\bf 211},	13-24.
\vsp
\rl  Miller, K.S. and  B. Ross (1993):
 {\it An Introduction to the Fractional
  Calculus and Fractional Differential Equations}
  (Wiley, New York).
\vsp
\rl Mitchell, B.J. (1995):
  Anelastic structure and evolution of the continental crust and
  upper mantle from seismic surface wave attenuation,
{\it Rev. Geophys.}, {\bf 33}, 441-462 (1995).
\vsp
\rl Murphy, W.F. (1982):
 Effect of partial water saturation on attenuation in sandstones,
{\it J. Acoust. Soc. Am.}, {\bf 71}, 1458-1468.
\vsp
\rl  Nigmatullin, R.R. (1986):
  The realization of the generalized transfer equation in a medium with
  fractal geometry,
   {\it  Phys. Stat. Sol. B}, {\bf 133}, 425-430.
  [English transl. from Russian]
\vsp
\rl O'Connell, R.J. and B. Budiansky (1978):
  Measures of dissipation in viscoelastic media,
  {\it Geophys. Res. Lett.}, {\bf 5}, 5-8.
\vsp
\rl Oldham, K.B. and  J. Spanier (1974):
 {\it The Fractional Calculus}
   (Academic Press, New  York).
\vsp
\rl Pipkin, A.C. (1986): {\it Lectures on Viscoelastic Theory}
  (Springer-Verlag, New York), Ch. 4, pp. 56-76. [1-st ed, 1972]
\vsp
\rl Ranalli, G. (1987): {\it Rheology of the Earth}
 (Allen \& Unwin, London).
\vsp
\rl Rossikhin, Yu. A. and M.V. Shitikova (1997):
Application of fractional calculus to dynamic problems of
linear and nonlinear hereditary mechanics of solids,
{\it Appl. Mech. Rev}, {\bf 50}, 15-67.
\vsp
\rl Sabadini, R., Yuen, D.A. and P. Gasperini (1985):
  The effects of transient rheology on the interpretation of lower
mantle viscosity, {\it Geophys. Res. Lett.}, {\bf 12}, 361-364.
\vsp
\rl Sabadini, R., Smith, B.K. and D.A. Yuen (1987):
  Consequences of experimental transient rheology,
{\it Geophys. Res. Lett.}, {\bf 14}, 816-819.
\vsp
\rl  Samko S.G.,  Kilbas, A.A.	and   O.I. Marichev (1993):
  {\it Fractional Integrals and Derivatives, Theory and Applications},
   (Gordon and Breach, Amsterdam).
   [Engl. Transl. from Russian, {\it Integrals and Derivatives of
    Fractional	 Order and Some of  Their Applications},
    Nauka i Tekhnika, Minsk (1987)]
\vsp
\rl Savage, J.C. and M.E. O'Neill (1975):
 The relation between the Lomnitz and the Futterman theories of
  internal friction,
{\it J. Geophys. Res.},   {\bf 80}, 249-251.
\vsp
\rl Schneider,	W.R. and W. Wyss (1989):
  Fractional diffusion and wave equations,
  {\it J. Math. Phys.},  {\bf 30}, 134-144.
\vsp
\rl Spencer , J. W. (1981):
  Stress relaxation at low frequencies in fluid saturated rocks;
  attenuation and modulus dispersion,
  {\it J. Geophys. Res.},   {\bf 86}, 1803-1812.
\vsp
\rl  Strick, E. (1967):
  The determination of $Q\,, $ dynamic viscosity and creep curves
  from wave propagation measurements,
  {\it Geophys. J. R. Astr. Soc.}, {\bf 13}, 197-218.
\vsp
\rl  Strick, E. (1970):
  A predicted pedestal effect for pulse propagation in
 constant-$Q$ solids,	{\it  Geophysics}, {\bf 35},   387-403.
\vsp
\rl  Strick, E. (1982):
   Application of linear viscoelasticity to seismic wave propagation,
   in {\it Wave Propagation in Viscoelastic Media}, edited by F. Mainardi
  (Pitman, London), pp. 169-193.  [Res. Notes in Maths, Vol. 52]
\vsp
\rl  Strick, E. and F. Mainardi (1982):
  On a general class of constant $Q$ solids,
  {\it Geophys. J. R. Astr. Soc.}, {\bf 69}, 415-429.
\vsp
\rl  Strick, E. (1984):
   Implication of Jeffreys-Lomnitz transient creep,
    {\it J. Geophys. Res.}, {\bf 89}, 437-451.
\vsp
\rl Yuen, D.A., Sabadini, R.,  Gasperini, P. and E. Boschi (1986):
  On transient rheology and glacial isostasy,
 {\it J. Geophys. Res.}, {\bf 91}, 11420-11438.
\vsp
\end{document}